\documentclass[a4paper,11pt]{article}
\usepackage{pos,xcolor}
\usepackage[vcentermath]{youngtab}

\usepackage{xcolor}

\newcommand{\ber}{\begin{eqnarray}}
\newcommand{\eer}{\end{eqnarray}}
\newcommand{\na}{\nabla}
\newcommand{\nn}{\nonumber}
\newcommand{\re}[1] {(\ref{#1})}

\title{Uses of Killing and Killing-Yano Tensors}
 \ShortTitle{UUITP-07/22}

\author*[a,b]{Ulf Lindstr\"om}
\author[b]{\"Ozg\"ur Sar\i o\u glu}

\affiliation[a]{Theoretical Physics, Dept. of Physics and Astronomy, Uppsala University,\\
  SE-751 20 Uppsala, Sweden}

\affiliation[b]{Department of Physics, Faculty of Arts and Sciences, Middle East Technical University,\\
06800, Ankara, Turkey}

\emailAdd{ulf.lindstrom@physics.uu.se}
\emailAdd{sarioglu@metu.edu.tr}

\abstract{In this contribution we have collected some facts about Killing and Killing-Yano tensors 
that we feel are of general interest for researchers working on problems that rely on differential 
geometry. We also include some of our recent studies pertaining to currents, charges 
and (super)invariants for particles and tensionless strings.}

\FullConference{%
  Corfu Summer Institute 2021 "School and Workshops on Elementary Particle Physics and Gravity"\\
  29 August - 9 October 2021\\
  Corfu, Greece
}

\tableofcontents

\begin{document}
\maketitle

\section{Introduction}
Killing tensors, Killing-Yano forms and Killing-Yano tensors have many uses: Killing tensors 
correspond to ``hidden'' symmetries of bosonic models 
\cite{Santillan:2011sh, Frolov:2008jr, Frolov:2017kze}.
They are also instrumental for separating variables in General Relativity \cite{Carter, 
Walker:1970un} and string equations \cite{Chervonyi:2015ima}, and are used in the study 
of G-structures \cite{Papadopoulos:2007gf, Papadopoulos:2011cz}. Killing tensors 
characterise the symmetries of Laplacians \cite{Eastwood:2002su}. Killing-Yano tensors square 
to Killing tensors and characterise the symmetries of the Dirac equation \cite{Cariglia:2012ci} and 
are also related to novel supersymmetries in sigma models and strings \cite{Gibbons:1993ap, 
DeJonghe:1996fb}. Supersymmetric Killing-Yano tensors \cite{Howe:2018lwu} characterise 
the symmetries of super Laplacians \cite{Howe:2016iqw, Kuzenko:2020www, Kuzenko:2019tys}. 
Finally, KTs arise in the context of 
hyperK\"ahler geometry \cite{Lindstrom:2009afn}. Of particular interest in the present 
context is the relation of Killing-Yano tensors to asymptotic conserved charges \cite{Kastor:2004jk, 
Olmez:2005by}. In this presentation we will give lightning reviews of a few of these topics and 
report on some new results \cite{Lindstrom:2021qrk, Lindstrom:2021dpm} and \cite{Lindstrom:2022iec}.

The paper is organised as follows: Section 2 contains the definition of Killing tensors and 
illustrations of their use in constructing invariants along geodesics, separating variables and their 
role in determining symmetries of the Laplacian. Section 3 contains the definition of Killing-Yano 
forms and illustrations of their use for constructing rank two Killing tensors, conserved currents,  
and asymptotic charges. In three dimensions we introduce a current based on the Cotton tensor 
and display its form in superspace as well as in ordinary space. Section 4 contains the definition 
of Killing-Yano tensors with both symmetric and antisymmetric sets of indices and an illustration 
of how they are used to find invariants of spinning particles and for spinning tensionless strings.

\section{Killing tensors}
In this section we introduce Killing tensors (KTs), their generalisation to conformal Killing 
tensors (CKTs) and illustrate their usefulness in a couple of examples.\\

An $n$\! th rank Killing tensor is a completely symmetric tensor $f^{\mu_1\dots \mu_n}$ that 
satisfies the equation
\ber\label{kill}
\na_{(\mu_1}f_{\mu_2\dots \mu_{n+1})}=0~,
\eer
whereas a conformal Killing tensor is trace-free and satisfies 
\ber
\na_{(\mu_1}f_{\mu_2\dots \mu_{n+1})}=ng_{(\mu_1\mu_2}\bar f_{\mu_3\dots \mu_{n+1})}~,
\eer
where the rank $n-1$ tensor $\bar f$ is determined by tracing both sides:
\ber
 \bar f_{\mu_1\dots \mu_{n-1}}= 
 \frac 1 {D+2(n-1)}
\na_\nu f^\nu_{~\mu_1\dots \mu_{n-1}}~.
\eer{}

\subsection{Uses of KTs}
In this subsection we indicate some of the most important uses of KTs.\\

\centerline{\em Conservation along geodesics}
\bigskip

Let  $x^\mu(\tau)$ be a geodesic, an over dot denote $\tau$-derivative and let $p^\mu :=  \dot x^\mu$ 
be the tangent vector to the geodesic. The covariant directional derivative along the geodesic is 
denoted
  \ber
  \frac D {d\tau}=p^\mu\na_\mu
  \eer
  and the geodesic equation is
  \ber\label{geo}
  \frac {D p^\mu} {d\tau}=0~.
  \eer
  If $f$ is an $n$th rank Killing tensor, the quantity
  \ber
  Q=f_{\mu_1\dots \mu_n}p^{\mu_1}\dots p^{\mu_n}
  \eer
  is then conserved along the geodesic \cite{Penrose:1986ca},
  \ber
  &&\frac D {d\tau}Q=p^\mu\na_\mu Q=p^\mu\na_\mu f_{\mu_1\dots \mu_n}p^{\mu_1}\dots p^{\mu_n}\\[1mm]\nn
  &&=\na_{(\mu} f_{\mu_1\dots \mu_n)}p^{\mu}p^{\mu_1}\dots p^{\mu_n}=0~,
  \eer
where we used \re{geo} and \re{kill}. Note that this is a purely geometric construction. Since the equations of motion of point-particles are geodesics it can be used to construct invariants for 
particles moving in geometries that allow KTs. It can also be extended to provide invariants for 
spinning particles \cite{Howe:2018lwu} and for tensionless strings \cite{Lindstrom:2022iec}.\\ 
  
\centerline {\em Separation of variables}
\bigskip
  
Here we briefly describe one of the most important uses of KTs, for separation of variables. The 
literature on this is vast, see, e.g., \cite{Santillan:2011sh, Chervonyi:2015ima}. Below we sketch 
the logic following the presentation in \cite{Chervonyi:2015ima}.
   
Let $S= W(X)+{\textstyle \frac 1 2}t\mu^2$ be Hamilton's principal function and $W$ his 
characteristic function, for a system with Hamiltonian $H={\textstyle \frac 1 2}pg^{-1}p$. The 
Hamilton-Jacobi equation 
\ber
H(X,\frac {\partial S}{\partial q},t)+\frac {\partial S}{\partial t}=0~
\eer
then reads
\ber\label{HJ}
g^{\mu\nu}\partial_\mu W \partial_\nu W+\mu^2=0~,
\eer
where $\mu$ is a constant. 
Assume that the coordinates $(X)$ can be divided into two distinct groups denoted by $(x)$ and 
$(y)$. If the following conditions are satisfied,
\ber
&&W = W_x(x_1, . . . x_k) + W_y(y_{k+1} . . . y_n)\\[1mm]\nn
&&g^{\mu\nu} = \frac {X^{\mu\nu}(x) +Y^{\mu\nu}(y)}{f_x-f_y } ~,\quad X^{y\nu}=0~,~~~Y^{x\nu}=0~,\\[1mm]\nn
&&
\partial_{y} f_{x}=0, \quad \partial_{x} f_{y}=0~,
\eer
then \re{HJ} separates as
\ber\label{sep}
X^{\mu\nu}\partial_\mu W \partial_\nu W+\mu^2f_x =-Y^{\mu\nu}\partial_\mu W \partial_\nu W+\mu^2f_y
\eer
with the left hand side a function of $x$ only and the right hand side a function of only $y$. 
Applying the usual argument that the two sides must be separately constant, we find the following
integral of the motion
\ber
I=X^{\mu\nu}\partial_\mu W_x \partial_\nu W_x+\mu^2f_x=\big(X^{\mu\nu}-g^{\mu\nu}f_x\big)\partial_\mu W_y \partial_\nu W_y~.
\eer
It can then be shown that
\ber
I=-\Big(\frac{f_yX^{\mu\nu}+f_xY^{\mu\nu}}{f_x-f_y}\Big)\partial_\mu W \partial_\nu W=:K^{\mu\nu}\partial_\mu W \partial_\nu W
\eer
with $K^{\mu\nu}$ a second rank KT. 

The key to the usefulness of this relation is that the converse is also true, as described in \cite{Chervonyi:2015ima}: Given a second rank KT, one can use it to find the separation \re{sep}. 
\bigskip

\centerline{\em Symmetries of the Laplacian}
\bigskip

A symmetry of the Laplacian $\Delta$ is a linear differential operator ${\cal D} \ne \Delta$ such that 
\cite{Eastwood:2002su}
\ber
\Delta {\cal D} = \delta \Delta
\eer
for some linear differential operator $\delta$. Any linear differential operator on a Riemannian 
manifold may be written in the form
\ber
D=V^{\mu\nu\dots \rho}\na_\mu\na_\nu\dots \na_\rho+ \makebox{lower order terms}~,
\eer
where $V^{\mu\nu\dots \rho}$ is symmetric in its indices. This tensor is called {\em the symbol}
of $D$. 

It is shown in \cite{Eastwood:2002su} that any symmetry ${\cal D}$ of the Laplacian on a Riemannian 
manifold is equivalent to one whose symbol is a CKT $f$:
\ber
{\cal D}=f^{\mu\nu\dots \rho}\na_\mu\na_\nu\dots \na_\rho+ \makebox{lower order terms}.
\eer
Similar results hold for the Dirac operator \cite{Cariglia:2012ci} and in superspace for 
super-Laplacians \cite{Howe:2016iqw}.
\bigskip

Having exemplified the uses of KTs, we now turn to Killing-Yano and conformal Killing-Yano 
forms\footnote{We refer to the totally antisymmetric versions as Killing-Yano forms and reserve 
the label Killing-Yano to tensors with mixed symmetries.}.

\section{Killing-Yano forms}
Killing-Yano forms (KYFs) are generalisations of KTs to antisymmetric covariant tensors.\\

 An $n~\!$th rank KYF is an $n$-form $k$ that satisfies
\ber
\na_{(\mu_1}k_{\mu_2)\dots \mu_{n+1}}=0~,
\eer
or, equivalently,
\ber\label{kyf}
\na_{\mu_1}k_{\mu_2\dots \mu_{n+1}}=\na_{[\mu_1}k_{\mu_2\dots \mu_{n+1}]}~.
\eer{}
A conformal Killing-Yano form (CKYF) satisfies
\ber\label{cKYT}
\na_{\mu_1}{\mathscr {k}}_{\mu_2\dots \mu_{n+1}}=\na_{[\mu_1}{\mathscr {k}}_{\mu_2\dots \mu_{n+1}]}
+\frac n {D-n+1}g_{\mu_1[\mu_2}\bar {\mathscr {k}}_{\mu_3\dots \mu_{n+1}]}~,
\eer
where 
\ber
\bar {\mathscr {k}}_{\mu_1\dots \mu_{n-1}}=\na_{\mu}{\mathscr {k}}^{\mu}_{~\mu_1\dots \mu_{n-1}}~.
\eer
Moreover, it is called {\em closed} if 
\ber\label{cckyf}
\na_{[\mu_1}{\mathscr {k}}_{\mu_2\dots \mu_{n+1}]}=0~.
\eer
  
\subsection{Uses of KYFs}
  
In this subsection we indicate some of the most important uses of KYFs.
\bigskip
  
\centerline{\em The square of a KYF is a KT}
\bigskip

In applications, it is often easier to find KYFs than KTs for a given geometry. It is then gratifying 
that KYFs square to second rank KTs:
   
Let $k_{\mu_1\dots \mu_n}$ be a  KYF.  Consider the second rank tensor
\ber\label{KYtoK}
f_{\mu\nu}=k_{\mu\mu_2\dots \mu_n}k_\nu^{~~\!\mu_2\dots \mu_n}~.
\eer
Then
\ber
&&\na_\sigma f_{\mu\nu}=
\na_{[\sigma}k_{\mu\mu_2\dots \mu_n]}
k_\nu^{~\mu_2\dots \mu_n}+
k_{\mu}^{~\mu_2\dots \mu_{n}}
\na_{[\sigma}k_{\nu\mu_2\dots \mu_n]}\\[1mm]\nn
&&\Rightarrow \na_{(\sigma} f_{\mu\nu)}=0~.
\eer
So $f_{\mu\nu}$ is a Killing tensor. This construction also produces rank 2 CKTs from two 
rank $n$ CKYFs \cite{Houri:2010fr}. Note that when the CKYFs are different, we must explicitly symmetrise the free indices in \re{KYtoK}.\\
   
\centerline{\em Conserved currents}
\bigskip
   
 A covariantly conserved antisymmetric rank $n$ tensor field  $J$ is equivalent to a co-closed 
 $n$-form. By the Poincar\'e lemma, this means that it is equal to the co-derivative of an 
 $(n + 1)$-form $\ell$ in a simply-connected open set
 \ber
 \na^\mu J_{\mu\dots \mu_n} = 0 \Rightarrow J_{\mu_1\dots \mu_n} =\na^\mu \ell _{\mu \mu_1\dots \mu_n}~.
 \eer
 This can be used to construct conserved charges for a given $J$. An interesting  example is the Kastor-Traschen $\mathbb{KT}$ current \cite{Kastor:2004jk}:
 \ber
 J^{\mu_{1} \dots \mu_{n}} =
- \frac{(n-1)}{4} \, R^{[\mu_{1} \mu_{2}}\,_{\rho\sigma} \, k^{\mu_{3} \dots \mu_{n}] \rho\sigma}
+ (-1)^{n+1} \, R_{\rho}\,^{[\mu_{1}} \, k^{\mu_{2} \dots \mu_{n}] \rho}
- \frac{1}{2 n} \, R \, k^{\mu_{1} \dots \mu_{n}} \,,
\eer
where $k$ is a KYF and the geometry is represented by the curvature tensor and its contractions. The covariant divergence of $J$ vanishes due to Bianchi identities and the properties \re{kyf} of $k$.

In \cite{Lindstrom:2021qrk} we show how the $\mathbb{KT}$ current may be rewritten in terms of the 
Weyl and Schouten tensors and how it has several separately conserved constituents. Interestingly 
only the full current seems to allow for Abbott-Deser (AD) charges, to which we now turn.\\

\centerline{\em AD charges}
\bigskip

In the spirit of \cite{Abbott:1981ff} one may construct asymptotic charges for the Kastor-Traschen 
$\mathbb{KT}$ current. If the metric has the asymptotic form
 \ber
 g_{\mu\nu}= \bar g_{\mu\nu}+  h_{\mu\nu}
 \eer
 and the background geometry defined by $\bar g_{\mu\nu}$ has a KYF $\bar k$, then the linearised 
 $\mathbb{KT}$ current
 \ber
 J^{(L)\mu_1\dots \mu_n} =\frac{(n-1)}{4} \,R_L^{[\mu_{1} \mu_{2}}\,_{\rho\sigma} \, \bar k^{\mu_{3} \dots \mu_{n}] \rho\sigma}
+ (-1)^{n+1} \, R_{L\rho}\,^{[\mu_{1}} \, \bar k^{\mu_{2} \dots \mu_{n}] \rho}
- \frac{1}{2 n} \, R_L \, \bar k^{\mu_{1} \dots \mu_{n}} \,
\eer
will be background conserved for certain geometries such as asymptotically flat or asymptotically 
AdS ones. For $n=2$, one finds
\ber
Q^{\mu\nu} \sim \int_{\Sigma} \, dS_{i} \, \sqrt{|\bar{\gamma}|} \, \bar{\ell}^{\mu\nu i} ~,
\eer
which then represents a conserved ``charge''. Note that this requires deriving the potential 
$\bar{\ell}^{\mu\nu\rho}$. This derivation prompted the development of some mathematical tools.\\

\centerline{\em Identities}
 \bigskip
The study of identities and integrability conditions for KTs, CKTs, KYFs and CKYFs has a long history:
\cite{Kashiwada:1968fva, Tachibana1, Tachibana2, Batista:2014fpa}. Here we relate 
some recent results along this line of investigations.
  
 In deriving the explicit relation
 $J^{(L)}_{\mu_1\dots \mu_n} =\bar \na^\mu \bar \ell _{\mu \mu_1\dots \mu_n}$ needed for 
 construction of the charge, we use
 \ber\label{KTid}
 \nabla_{\mu} \nabla_{\nu} k_{\rho_{1} \dots \rho_{n}} = (-1)^{n+1} \frac{(n+1)}{2} \,
 R^{\sigma}\,_{\mu[\nu\rho_{1}} \, k_{\rho_{2} \dots \rho_{n}] \sigma} \,, 
 \eer
 which generalises the Killing vector relation 
 \( \nabla_{\mu} \nabla_{\nu} f_{\rho} = R^{\sigma}\,_{\mu\nu\rho} \, f_{\sigma} \).
 From this, we derive a number of new relations such as
 \ber
 R_{\mu\nu} \, k^{\mu\sigma} + R^{\mu\sigma} \, k_{\mu\nu} = 0 
 \eer
 for a rank 2 KYF. The corresponding identity for a general rank KYF $k$ reads
 \small
 \ber
R^{\mu}\,_{\nu} \, k_{[\sigma_{1} \dots \sigma_{n-1}] \mu} + (-1)^{n} \, R^{\mu}\,_{[\sigma_1} \, k_{\sigma_{2} \dots \sigma_{n-1}] \nu\mu} 
\qquad \qquad \qquad \qquad \qquad \qquad & 
\nonumber \\
+ (n-2) \left( (-1)^{n} R_{\nu\mu \lambda[\sigma_{1}} \, k_{\sigma_{2} \dots \sigma_{n-1}]}\,^{\lambda\mu}
    + \frac{1}{2} \, R_{\lambda\mu[\sigma_{1} \sigma_{2}} \, k_{\sigma_{3} \dots \sigma_{n-1}]\nu}\,^{\lambda\mu} \right) & = 0 \,. 
\eer
Several other new identities are to be found in \cite{Lindstrom:2021qrk}. The following one, involving
 the Einstein tensor 
 \ber
 G_{\mu\nu}=R_{\mu\nu}-{\textstyle \frac 1 2} g_{\mu\nu} R,
 \eer
 reads
\ber\label{fg}
k^{\mu\nu} \nabla_{\mu} G_{\nu\rho} = 0 ~.
\eer
The corresponding identity for a general rank KYF $k$ reads
\ber
(n-1) \left( \nabla^{\nu} \, R^{\mu}\,_{[\sigma_{1}} \right) k_{\sigma_{2} \dots \sigma_{n-1}] \mu\nu} 
+ \frac{1}{2} \, \left( \nabla^{\mu} R \right) k_{\mu [\sigma_{1} \dots \sigma_{n-1}]} = 0 \,.
\label{noEin}
\eer{}
We use \re{fg}  to prove that 
\ber
 K^{\mu\nu} =2 \, G_{\rho}\,^{[\mu} \, k^{\nu]\rho}
 \eer
 is a conserved ``current''. Note that in a given geometry this leads to a relation between the KYF and 
 the energy-momentum tensor on shell. More details may be found in  \cite{Lindstrom:2021qrk}.\\

The identity \re{KTid} may be generalized to CKYTs \re{cKYT} (as well as to geometries with 
torsion, see \cite{Lindstrom:2021dpm}). For a second rank CKYT $ \mathscr{k}$ , it reads
\ber
\nabla_{\mu} \nabla_{\nu} \mathscr{k}_{\rho\sigma} = 
- \frac{3}{2} R^{\tau}{}_{\mu[\nu\rho} \mathscr{k}_{\sigma]\tau} 
- \frac{3}{D-1} g_{\mu[\nu} \nabla_{\rho} \bar{ \mathscr{k}}_{\sigma]}
+ \frac{2}{D-1} \nabla_{\mu} \left( g_{\nu[\rho} \bar{ \mathscr{k}}_{\sigma]} \right) \,. 
\eer
 
\bigskip
 \centerline{\em PCCKYF}
 \bigskip
 
The closed conformal Killing-Yano forms defined in \re{cckyf} play an important role when $n=2$ 
and it is non-degenerate as a matrix. It is then called a {\em principal closed conformal Killing-Yano 
tensor} and is the starting point for constructing a hierarchy of KYTs and KTs that can be used to 
characterise the solution. E.g., for the Kerr-NUT-(A)dS family this principal tensor generates a  
hierarchy of Killing vectors that ensures complete integrability of geodesic motion and separability 
of the Hamilton-Jacobi, Klein-Gordon, and Dirac equations, see \cite{Frolov:2017kze}.
 
 \centerline{\em Cotton current comments}
 \bigskip
 
Above we have focused on the $\mathbb{KT}$ current.  Another current, the Killing-Yano Cotton 
current, was recently constructed in \cite{Lindstrom:2021dpm}. Here we briefly describe this current.
\bigskip
 
The Cotton tensor is defined in $D \geq 3$ dimensions as  
\ber
C_{\mu \nu\rho} \equiv 2 (D-2) \nabla _{[\rho} S_{\nu]\mu} =
2 \nabla _{[\rho} R_{\nu]\mu} - \frac{1}{(D-1)} g_{\mu[\nu} \nabla _{\rho]} R \,.
\eer
It satisfies 
\ber\nn
&& C_{\mu\nu\rho} = C_{\mu[\nu\rho]} \,, \\[1mm] \nn
&& C_{[\mu\nu\rho]} = 0 \,, \\[1mm]
&& \na^\mu C_{\mu\nu\rho} = 0~.
\eer
Recall that a second rank CKYF ${\mathscr {k}}$ satisfies
\ber
\na_{\mu}{\mathscr {k}}_{\nu\rho}=\na_{[\mu}{\mathscr {k}}_{\nu\rho]}
+\frac 2 {D-1}g_{\mu[\nu}\bar {\mathscr {k}}_{\rho]}~,
\eer 
and consider 
\ber
 J^\mu \equiv C^{\mu\nu\rho} {\mathscr {k}}_{\nu\rho} \,.
\eer
That this is a conserved current follows from
\ber
&&\na_\mu J^\mu = (\na_\mu C^{\mu\nu\rho}) {\mathscr {k}}_{\nu\rho}+ C^{\mu\nu\rho} \na_\mu {\mathscr {k}}_{\nu\rho} \\[1mm]\nn
&&= 0 + C^{\mu\nu\rho} (\na_{[\mu} {\mathscr {k}}_{\nu\rho]}+\frac{2}{D-1} \,  g_{\mu[\nu} \, \bar{{\mathscr {k}}}_{\rho]}) = 0~,
\eer
since $C$ is conserved, symmetric and traceless. We define a charge for this current. 
Since
\ber
\na_{\mu} J^{\mu} = \frac{1}{\sqrt{|g|}} \partial_{\mu} \left( \sqrt{|g|} \, J^{\mu} \right) = 0 \,,
\eer
one can define a conserved charge ${\cal Q}$ as
\ber
{\cal Q} \equiv \int_{\Sigma_{\tau}} d^{(D-1)} x \, J^{\mu} \, n_{\mu} \,,
\eer
with $n^{\mu}$ normal to the spacelike surface $\Sigma$. We apply this construction to the 
Pleba\'nski-Demia\'nski metric where we can carry out the integration for certain values of the 
metric parameters in \cite{Lindstrom:2021dpm}.\\

\centerline {$3D$}
\bigskip

In three dimensional topologically massive gravity, the Cotton tensor, or its descendant, the 
York tensor
\ber
C^{\mu\nu} = \frac{1}{\sqrt{|g|}} \epsilon^{\mu\sigma\rho} \nabla_{\sigma} S_{\rho}{}^{\nu} \,,
\eer
has a prominent role. In \cite{Deser:2003vh}, it is used to define asymptotic charges using a current 
based on the energy-momentum tensor and an asymptotic Killing vector. In \cite{Lindstrom:2021dpm}, 
we compare it to our Cotton current in $3D$ with some interesting results.
\bigskip

$3D$ is also our starting point for lifting conserved currents to supergravity to which we now turn.\\

\centerline{$3D~$\em conformal supergravity} 
\bigskip

In this section we use Greek letters for spinor indices while vector indices are represented by 
Latin letters or, equivalently, by pairs of spinor indices.

In three dimensions, conformal supergravity may be defined by the following algebra of 
covariant derivatives \cite{Butter:2013goa, Kuzenko:2012ew} 
\ber\nn
&&\{\na_\alpha,\na_\beta\}=2i\na_{\alpha\beta}~,\\[1mm]\nn
&&[\na_a,\na_\alpha]={\textstyle \frac 14}(\gamma_a)_{\alpha}^{~~\beta}W_{\beta\gamma\delta}K^{\gamma\delta}~,\\[1mm]
&&[\na_a,\na_b]=-{\textstyle \frac i8}\epsilon_{abc}(\gamma^c)^{\alpha\beta}\na_\alpha W_{\beta\gamma\delta}K^{\gamma\delta}
-{\textstyle \frac 14}\epsilon_{abc}(\gamma^c)^{\alpha\beta}W_{\alpha\beta\gamma}S^\gamma~.
\eer 
The notation is that spinor indices are $\alpha, \beta, \dots$, vector tangent space indices are $a,b,
\dots$. The usual convention that a vector index is represented by a symmetric pair of spinor indices 
also applies. Thus $K_{\alpha\beta}$ is the vector generator of special conformal transformations while 
$S^\gamma$ generates $S$ supersymmetry transformations.

The super Cotton tensor \cite{Kuzenko:2012ew} $W$ obeys
\ber\nn
&&W_{\alpha\beta\gamma}=W_{(\alpha\beta\gamma)}\\[1mm]\nn
&&\na^\alpha W_{\alpha\beta\gamma}=0\\[1mm]
&&K_aW_{\alpha\beta\gamma}=0~,
\eer
where the last relation identifies $W$ as a primary field.
\bigskip

\centerline{\em The super Cotton current}
\bigskip

Armed with these relations, we turn to the super Cotton current \cite{Lindstrom:2021dpm}.\\

We take a superconformal Killing supervector field $\xi$ to be given by
\ber
\xi=\xi^a\na_a+\xi^\alpha\na_\alpha~,
\eer
with $\xi^a$ a primary field. It follows that
\ber\nn
&&\na_{(a}\xi_{b)}={\textstyle \frac 1 3}\eta_{ab}\na_c\xi^c\\[1mm]
&&\na^{\beta\gamma}\xi^\alpha=-{\textstyle \frac 2 3}\epsilon^{\alpha(\beta}\na^{\gamma)}_{~\sigma}\xi^\sigma~,
\eer
where the next to last relation defines a conformal Killing vector and the last one a conformal 
Killing spinor.

We may now construct a supergravity version of our Cotton current. To this end, we define
\ber\nn
&&k_\alpha=W_{\alpha\beta\gamma}\xi^{\beta\gamma}\\[1mm]
&&k_{\alpha\beta}=\na_\alpha k_\beta=(\na_{(\alpha} W_{\beta)\gamma\delta})\xi^{\gamma\delta}+4iW_{\alpha\beta\gamma}\xi^{\gamma}~.
\eer{}
These satisfy
\ber\nn
&&\na^\alpha k_\alpha=0\\[1mm]
&&\na^{\alpha\beta}k_{\alpha\beta}=0~,
\eer
and the lowest component of the first part of $k_{\alpha\beta}$ is the bosonic Cotton current.
In a more covariant form we have
\ber
(k^A)=(k^{\alpha\beta},k^\alpha)
\eer
with
\ber
\na_Ak^A=0~.
\eer

We note that this construction opens the novel field of super invariants from super Killing tensors.

\section{Killing-Yano tensors}

In this section we introduce the mixed KYTs and exemplify some of their uses. Here we only consider 
a Minkowski space background. 

A product of a conformal Killing vector ${\mathscr {k}}_a$ and a CKY 2-form ${\mathscr {k}}_{\mu\nu}$ 
projected onto the highest weight representation gives an object
\ber
A_{\mu,\nu\rho}:={\mathscr {k}}_{\mu(\nu} {\mathscr {k}}_{\rho)}+\frac{1}{(n-1)}\left(\eta_{\mu(\nu} ({\mathscr {k}}\cdot {\mathscr {k}})_{\rho)}-\eta_{\nu\rho} ({\mathscr {k}}\cdot {\mathscr {k}})_\mu\right)~,
\label{2.1}
\eer
In \cite{Howe:2018lwu} this construction is generalised\footnote{There is a different generalisation of 
KYFs in \cite{Kress}.} to tensors $A_{p_1,p_2,\ldots ,q}$ with $p_1\geq p_2\geq \ldots$ being the 
number of boxes minus one in the columns starting from the left. For example, 
\ber
A_{p_1,p_2,q}\sim\ \ \ \ \ \overbrace{\young(\hfil\hfil\hfil\hfil\hfil,\hfil\hfil,\hfil\hfil,\hfil,\hfil)}^{q}~,
\label{2.5}
\eer
where there are $p_1+1$ boxes in the first column and $p_2+1$ in the second. The differential 
constraint satisfied by this tensor is
\ber
\partial A_{p_1,p_2,q}\ni \  \overbrace{\young(\hfil\hfil\hfil\hfil\hfil\hfil,\hfil\hfil,\hfil\hfil,\hfil,\hfil)}^{q+1}
= 0\ .
\label{2.5.1}
\eer
For the particular case $A_{p,q}$ the Young tableau is
\ber
A_{p,q}\sim\ \ \ \ \ {\overbrace{\young(\hfil\hfil\hfil\hfil\hfil,\hfil,\hfil,\hfil)}^{q}}
\label{2.2}
\eer
with $(p+1)$ boxes in the first column. The differential constraint satisfied by such a CKYT is that, 
when a derivative is applied to $A_{p,q}$, the traceless tensor corresponding to the Young tableau 
with one extra box on the first row has to vanish, i.e.,
\ber
\partial A_{p,q}\ni \overbrace{\young(\hfil\hfil\hfil\hfil\hfil,\hfil,\hfil,\hfil)}^{q+1}\ = 0\ .
\label{2.2.1}
\eer
This kind of tensors  appears naturally in the context of the spinning particle \cite{Gibbons:1993ap, 
vanHolten:1992bu}.

\subsection{Uses of KYTs}
{These Killing-Yano tensors can, e.g., be used for finding invariants of spinning particles and for
spinning tensionless strings.}

\centerline{\em Spinning Particle}
\bigskip

In \cite{Lindstrom:2021dpm} invariants for the spinning particle are constructed using $ A_{p,q}$ 
CKYTs\footnote{Integrability for the spinning particle is discussed in a different context in \cite{Kubiznak:2011ay}.}. It is shown that invariants take the form
\ber
F=\lambda^p A_{p,q}p^q+\alpha(p,q)\lambda^{p+2} dA_{p+2,q-1} p^{q-1}:=A+dA \,,
\eer
where $A_{p,q}$ is in the representation \re{2.2}, satisfies the constraint \re{2.2.1} and
\ber
\alpha(p,q):= i\frac{(-1)^{(p+1)} q}{(1+p+q)}\ ,
\eer{}
\ber
\left(dA_{p+2,q-1}\right)_{\nu_1\ldots \nu_{p+2},\mu_1\ldots \mu_{q-1}}
:=\partial_{[\nu_1}A_{\nu_2\ldots \nu_{p+1},\nu_{p+2}]\mu_1\ldots \mu_{q-1}}~.\ 
\eer{}

\centerline{\em Spinning tensionless string}
\bigskip

In \cite{Lindstrom:2022iec} we take advantage of the close relation between the spinning 
tensionless string \cite{Lindstrom:1990ar} and the spinning particle \cite{Howe:1988ft}: In a 
particular gauge the spinning tensionless string is a collection of spinning particles obeying 
certain constraints. We are then able to generalise the construction of invariants for the spinning 
particle to the spinning tensionless string.
 
\section{Summary}
We have given the definitions of KTs, CKTs, KYFs, CKYFs and KYTs and listed a number 
of applications for each of these. In addition, we have presented some recent results on 
conserved currents, (asymptotic) charges, nontrivial identities and their applications to spinning 
particles and spinning tensionless strings. In particular, the extension of conserved currents to 
$3D$ supergravity represents a novel and promising line of research.

\bigskip
\noindent{\bf Acknowledgments}\\
We thank Sergei Kuzenko for commenting on the supergravity part and David Kubiznak for comments and references. The research of 
U.L. is supported in part by the 2236 Co-Funded Scheme2 (CoCirculation2) of T\"UB{\.I}TAK 
(Project No:120C067)\footnote{\tiny However the entire responsibility for the publication is ours. 
The financial support received from T\"UB{\.I}TAK does not mean that the content of the 
publication is approved in a scientific sense by T\"UB{\.I}TAK.}.\\


\begin{thebibliography}{99}
\bibitem{Santillan:2011sh}
O.~P.~Santillan,
``Hidden symmetries and supergravity solutions,''
\href{https://doi.org/10.1063/1.3698087}{{\em J.\ Math.\ Phys.} {\bfseries 53} (2012) 043509};
\href{http://arxiv.org/abs/1108.0149}{{\ttfamily arXiv:1108.0149 [hep-th]}}.
  
\bibitem{Frolov:2008jr}
V.~P.~Frolov and D.~Kubiznak,
``Higher-Dimensional Black Holes: Hidden Symmetries and Separation of Variables,''
\href{http://dx.doi.org/10.1088/0264-9381/25/15/154005}{{\em Class. Quant. Grav.} {\bfseries 25}
154005 (2008)};
 \href{http://arxiv.org/abs/0802.0322}{{\ttfamily arXiv:0802.0322 [hep-th]}}.

\bibitem{Frolov:2017kze}
V.~Frolov, P.~Krtous and D.~Kubiznak,
``Black holes, hidden symmetries, and complete integrability,''
\href{http://dx.doi.org/10.1007/s41114-017-0009-9}{{\em Living\ Rev.\ Rel.} {\bfseries 20}, 
no.1, 6 (2017)};
 \href{http://arxiv.org/abs/1705.05482}{{\ttfamily arXiv:1705.05482 [gr-qc]}}.
  
 \bibitem{Carter}
 B.~Carter, 
 ``Global Structure of the Kerr Family of Gravitational Fields,"
 \href{https://doi.org/10.1103/PhysRev.174.1559}{{\em Phys. Rev.} {\bfseries 174} (1968) 1559}.
 
\bibitem{Walker:1970un}
M.~Walker and R.~Penrose,
``On quadratic first integrals of the geodesic equations for type \{22\} spacetimes,''
\href{https://doi.org/10.1007/BF01649445}{{\em Commun. Math. Phys.} {\bfseries 18} (1970) 
265-274}.

\bibitem{Chervonyi:2015ima}
Y.~Chervonyi and O.~Lunin,
``Killing(-Yano) Tensors in String Theory,''
\href{http://dx.doi.org/10.1007/JHEP09(2015)182}{{\em JHEP} {\bfseries 09} (2015) 182};
\href{http://arxiv.org/abs/1505.06154}{{\ttfamily arXiv:1505.06154 [hep-th]}}.


\bibitem{Papadopoulos:2007gf}
G.~Papadopoulos,
``Killing-Yano equations and G-structures,''
\href{http://dx.doi.org/10.1088/0264-9381/25/10/105016}{{\em Class.\ Quant.\ Grav.} {\bfseries 25} (2008) 105016};
\href{http://arxiv.org/abs/0712.0542}{{\ttfamily arXiv:0712.0542 [hep-th]}}.
  
\bibitem{Papadopoulos:2011cz}
G.~Papadopoulos,
``Killing-Yano equations with torsion, worldline actions and G-structures,''
\href{http://dx.doi.org/10.1088/0264-9381/29/11/115008}{{\em Class.\ Quant.\ Grav.} {\bfseries 29} (2012) 115008};
\href{http://arxiv.org/abs/1111.6744}{{\ttfamily arXiv:1111.6744 [hep-th]}}.
  
\bibitem{Eastwood:2002su}
M.~G.~Eastwood,
``Higher symmetries of the Laplacian,''
\href{http://dx.doi.org/10.4007/annals.2005.161.1645}{{\em Annals\ Math.} {\bfseries 161} (2005), 1645-1665};
\href{http://arxiv.org/abs/hep-th/0206233}{{\ttfamily arXiv:hep-th/0206233 [hep-th]}}.

\bibitem{Cariglia:2012ci}
M.~Cariglia,
``Hidden Symmetries of the Dirac Equation in Curved Space-Time,''
\href{https://doi.org/10.1007/978-3-319-06761-2_4}
{{\em In: Bičák J., Ledvinka T. (eds) 
Relativity and Gravitation. Springer\ Proceedings\ in}{\em\ Physics,\ vol\ 157.\ Springer,\ Cham.} \\
{\bfseries 157} (2014) 25}; 
\href{http://arxiv.org/abs/1209.6406}{{\ttfamily arXiv:1209.6406 [gr-qc]}}.

\bibitem{Gibbons:1993ap}
G.~W.~Gibbons, R.~H.~Rietdijk and J.~W.~van Holten,
``Susy in the sky,''
\href{http://dx.doi.org/10.1016/0550-3213(93)90472-2}{{\em Nucl. Phys. B} {\bfseries 404} (1993) 42-64};
\href{http://arxiv.org/abs/hep-th/9303112}{{\ttfamily arXiv:hep-th/9303112 [hep-th]}}.

\bibitem{DeJonghe:1996fb}
F.~De Jonghe, K.~Peeters and K.~Sfetsos,
``Killing-Yano supersymmetry in string theory,''
\href{http://dx.doi.org/10.1088/0264-9381/14/1/007}{{\em Class. Quant. Grav.} {\bfseries 14} (1997) 35-46};
\href{http://arxiv.org/abs/hep-th/9607203}{{\ttfamily arXiv:hep-th/9607203 [hep-th]}}.

\bibitem{Howe:2018lwu}
P.~S.~Howe and U.~Lindstr\"om,
``Some remarks on (super)-conformal Killing-Yano tensors,''
\href{https://doi.org/10.1007/JHEP11(2018)049}{{\em JHEP} {\bfseries 11} (2018) 049};
\href{http://arxiv.org/abs/1808.00583}{{\ttfamily arXiv:1808.00583 [hep-th]}}.

\bibitem{Howe:2016iqw}
P.~S.~Howe and U.~Lindstr\"om,
``Super-Laplacians and their symmetries,''
\href{https://dx.doi.org/10.1007/JHEP05(2017)119}{{\em JHEP} {\bfseries 05} (2017), 119};
\href{http://arxiv.org/abs/1612.06787}{{\ttfamily arXiv:1612.06787 [hep-th]}}.

\bibitem{Kuzenko:2020www}
S.~M.~Kuzenko, U.~Lindstr\"om, E.~S.~N.~Raptakis and G.~Tartaglino-Mazzucchelli,
``Symmetries of $ \mathcal{N} $ = (1, 0) supergravity backgrounds in six dimensions,''
\href{https://dx.doi.org/10.1007/JHEP03(2021)157}{{\em JHEP} {\bfseries 03} (2021), 157};
\href{http://arxiv.org/abs/2012.08159}{{\ttfamily arXiv:2012.08159 [hep-th]}}.

\bibitem{Kuzenko:2019tys}
S.~M.~Kuzenko and E.~S.~N.~Raptakis,
``Symmetries of supergravity backgrounds and supersymmetric field theory,''
\href{https://dx.doi.org/10.1007/JHEP04(2020)133}{{\em JHEP} {\bfseries 04} (2020), 133};
\href{http://arxiv.org/abs/1912.08552}{{\ttfamily arXiv:1912.08552 [hep-th]}}.

\bibitem{Lindstrom:2009afn}
U.~Lindstr\"om and M.~Ro\v cek,
``Properties of hyperk\"ahler manifolds and their twistor spaces,''
\href{https://dx.doi.org/10.1007/s00220-009-0923-0}{{\em Commun.\ Math.\ Phys.} {\bfseries 293} 
(2010) 257};
\href{http://arxiv.org/abs/0807.1366}{{\ttfamily arXiv:0807.1366 [hep-th]}}.

\bibitem{Kastor:2004jk}
D.~Kastor and J.~Traschen,
``Conserved gravitational charges from Yano tensors,''
\href{http://dx.doi.org/10.1088/1126-6708/2004/08/045}{{\em JHEP} {\bfseries 08} (2004) 045};
\href{http://arxiv.org/abs/hep-th/0406052}{{\ttfamily arXiv:hep-th/0406052 [hep-th]}}.

\bibitem{Olmez:2005by}
S.~\"Olmez, \"O.~Sar\i{}o\u{g}lu, and B.~Tekin,
``Mass and angular momentum of asymptotically ads or flat solutions in the topologically massive gravity,''
\href{http://dx.doi.org/10.1088/0264-9381/22/20/014}{{\em Class. Quant. Grav.} {\bfseries 22}, 4355-4362 (2005)};
\href{http://arxiv.org/abs/gr-qc/0507003}{{\ttfamily arXiv:gr-qc/0507003 [gr-qc]}}.

\bibitem{Lindstrom:2021qrk}
U.~Lindstr\"om and \"O.~Sar\i{}o\u{g}lu,
``New currents with Killing-Yano tensors,''
\href{http://dx.doi.org/10.1088/1361-6382/ac1871}{{\em Class. Quant. Grav.} {\bfseries 38} 
no.19, 195011 (2021)};
 \href{http://arxiv.org/abs/2104.12451}{{\ttfamily arXiv:2104.12451 [hep-th]}}.

\bibitem{Lindstrom:2021dpm}
U.~Lindstr\"om and \"O.~Sar\i{}o\u{g}lu,
``Killing-Yano Cotton Currents,''
 \href{http://arxiv.org/abs/2110.03470}{{\ttfamily arXiv:2110.03470 [hep-th]}}.
Accepted for publication in JHEP.

\bibitem{Lindstrom:2022iec}
U.~Lindstr\"om and \"O.~Sar\i{}o\u{g}lu,
``Tensionless Strings and Killing(-Yano) Tensors,''
 \href{http://arxiv.org/abs/2202.06542}{{\ttfamily arXiv:2202.06542 [hep-th]}}.

\bibitem{Penrose:1986ca}
R.~Penrose and W.~Rindler,
\href{https://dx.doi.org/10.1017/CBO9780511524486}{{\ttfamily SPINORS AND SPACE-TIME, 
VOL. 2:}} SPINOR AND TWISTOR METHODS IN SPACE-TIME GEOMETRY, Cambridge 
Univ. Press (1986).

\bibitem{Houri:2010fr}
T.~Houri, D.~Kubiznak, C.~M.~Warnick and Y.~Yasui,
``Generalized hidden symmetries and the Kerr-Sen black hole,''
\href{https://doi.org/10.1007/JHEP07(2010)055}{{\em JHEP} {\bfseries 07} (2010), 055};
 \href{http://arxiv.org/abs/1004.1032}{{\ttfamily arXiv:1004.1032 [hep-th]}}.

\bibitem{Abbott:1981ff}
L.~F.~Abbott and S.~Deser,
``Stability of Gravity with a Cosmological Constant,''
\href{http://dx.doi.org/10.1016/0550-3213(82)90049-9}{{\em Nucl. Phys. B} {\bfseries 195} (1982) 76-96}.

\bibitem{Kashiwada:1968fva}
T.~Kashiwada,
``On conformal Killing tensor,''
{{\em Nat.\ Sci.\ Rep.\ Ochanomizu Univ.} {\bfseries 19} (1968) no.2, 67-74}.

\bibitem{Tachibana1}
S.~Tachibana and T.~Kashiwada,
``On the integrability of Killing-Yano's equation,'' 
{{\em J.\ Math.\ Soc.\ Japan} {\bfseries 21} (1969) No.2, 259-265}.

\bibitem{Tachibana2}
S.~Tachibana,
``On conformal Killing tensor in a Riemannian space'', 
{{\em T\^ohoku\ Math.\ Journ.} {\bfseries 21} (1969), 56-64}.

\bibitem{Batista:2014fpa}
C.~Batista,
``Integrability Conditions for Killing-Yano Tensors and Conformal Killing-Yano Tensors,''
\href{https://doi.org/10.1103/PhysRevD.91.024013}{{\em Phys.\ Rev.\ D}{\bfseries 91}, 024013 
 (2015)};
 \href{http://arxiv.org/abs/1406.3069}{{\ttfamily arXiv:1406.3069 [gr-qc]}}.

\bibitem{Deser:2003vh}
S.~Deser and B.~Tekin,
``Energy in topologically massive gravity,''
\href{http://dx.doi.org/10.1088/0264-9381/20/21/L01}{{\em Class. Quant. Grav.} {\bfseries 20}, 
L259 (2003)};
\href{http://arxiv.org/abs/gr-qc/0307073}{{\ttfamily arXiv:gr-qc/0307073 [gr-qc]}}.

\bibitem{Butter:2013goa}
D.~Butter, S.~M.~Kuzenko, J.~Novak and G.~Tartaglino-Mazzucchelli,
``Conformal supergravity in three dimensions: New off-shell formulation,''
\href{https://doi.org/10.1007/JHEP09(2013)072}{{\em JHEP} {\bfseries 09} (2013), 072};
\href{https://arxiv.org/abs/1305.3132}{{\ttfamily arXiv:1305.3132 [hep-th]}}.

\bibitem{Kuzenko:2012ew}
S.~M.~Kuzenko and G.~Tartaglino-Mazzucchelli,
``Conformal supergravities as Chern-Simons theories revisited,''
\href{https://dx.doi.org/10.1007/JHEP03(2013)113}{{\em JHEP} {\bfseries 03} (2013), 113};
 \href{http://arxiv.org/abs/1212.6852}{{\ttfamily arXiv:1212.6852 [hep-th]}}.

\bibitem{Kress}
J.~ Kress,
``Generalised Conformal Killing-Yano Tensors: Applications to Electrodynamics,''
\href{https://web.maths.unsw.edu.au/~jonathan/thesis/thesis.pdf} {{\ttfamily Doctoral Thesis}}, 
University of Newcastle, Dept. of Math. (1997).  

\bibitem{vanHolten:1992bu}
  J.~W.~van Holten and R.~H.~Rietdijk,
  ``Symmetries and motions in manifolds,''
  \href{https://dx.doi.org/10.1016/0393-0440(93)90079-T}{{\em J.\ Geom.\ Phys.} {\bfseries 11} (1993) 559};
\href{http://arxiv.org/abs/hep-th/9205074}{{\ttfamily arxiv:hep-th/9205074}}.
  
  \bibitem{Kubiznak:2011ay}
D.~Kubiznak and M.~Cariglia,
``Integrability of spinning particle motion in higher-dimensional black hole spacetimes,''
\href{https://doi.org/10.1103/PhysRevLett.108.051104}{{\em Phys. Rev. Lett.} {\bfseries 108}, 051104 (2012)};
\href{http://arxiv.org/abs/1110.0495}{{\ttfamily arXiv:1110.0495 [hep-th]}}.
  
\bibitem{Lindstrom:1990ar}
U.~Lindstr\"om, B.~Sundborg and G.~Theodoridis,
``The zero tension limit of the spinning string,''
\href{https://doi.org/10.1016/0370-2693(91)91094-C}{{\em Phys. Lett. B} {\bfseries 258} (1991) 331-334}.

\bibitem{Howe:1988ft}
P.~S.~Howe, S.~Penati, M.~Pernici and P.~K.~Townsend,
``Wave Equations for Arbitrary Spin From Quantization of the Extended Supersymmetric Spinning 
Particle,''
\href{https://dx.doi.org/10.1016/0370-2693(88)91358-5}{{\em Phys. Lett. B} {\bfseries 215} (1988) 
555-558}.

\end{thebibliography}
\end{document}